# Data Aggregation, Fusion and Recommendations for Strengthening Citizens Energy-aware Behavioural Profiles

Eleni Fotopoulou, Anastasios Zafeiropoulos
R&D Department,
Ubitech Ltd.
Athens, Greece

Fernando Terroso, Aurora Gonzalez, Antonio Skarmeta
Dept. Ingeniería de la Información y las Comunicaciones, Facultad de Informática, Universidad de Murcia
Murcia, Spain

Umutcan Şimşek, Anna Fensel
Semantic Technology Institute
Innsbruck, Austria

*Abstract*— In this paper, ENTROPY platform, an IT ecosystem for supporting energy efficiency in buildings through behavioural change of the occupants is provided. The ENTROPY platform targets at providing a set of mechanisms for accelerating the adoption of energy efficient practices through the increase of the energy awareness and energy saving potential of the occupants. The platform takes advantage of novel sensor networking technologies for supporting efficient sensor data aggregation mechanisms, semantic web technologies for unified data representation, machine learning mechanisms for getting insights from the available data and recommendation mechanisms for providing personalised content to end users. These technologies are combined and provided through an integrated platform, targeting at leading to occupants' behavioural change with regards to their energy consumption profiles.

## I. Introduction

Buildings are estimated to be responsible for 40% of energy consumption and 36% of $CO_2$ emissions in the EU [1]. By improving the energy efficiency of buildings, it is estimated that the total EU energy consumption can be reduced by 5% to 6%, while $CO_2$ emissions can be reduced by about 5%. Towards this direction, the role of Information and Communication Technologies (ICT) is considered crucial, given that -as stated in the Smarter2030 report [2], ICT has the potential to enable a 20% reduction of global $CO_2$ equivalent emissions by 2030. Furthermore, it is stated that the emissions avoided through the use of ICT are nearly ten times greater than the emissions generated by deploying it [2], minimizing any concerns with regards to the energy consumption introduced via the deployment of the ICT equipment. A report from European Environmental Agency [4] shows that measures targeting behavioural change of consumers may help to achieve energy savings up to 20%, which would help member states significantly, to achieve the goal of reducing the primary energy consumption by 20%.

In the current manuscript, an IT ecosystem is presented aiming at exploiting the potential provided by a set of novel ICT technologies for supporting occupants behavioural change through energy consumption awareness. These technologies include Internet of Things (IoT) technologies, sensor data aggregation and fusion mechanisms including mobile crowdsensing, personalised recommendation mechanisms and machine learning techniques. The provided platform, as it is developed within the framework of the ENTROPY H2020 project [3], provides a set of integrated and well-interconnected tools for enabling the design, development and provision of personalised services, mobile applications and serious games targeted to buildings' occupants.

The structure of the manuscript is as follows: in section two, the ENTROPY platform overall architecture is detailed, focusing mainly on the data aggregation mechanisms, data fusion mechanisms and the recommendations and analytics framework; in section three, a short overview of the current status of the implementation of the platform as well as the targeted use cases are described, while section four provides main conclusions and details for planned future work towards the final release of the platform.

## II. Conceptual Architecture

The ENTROPY architecture follows a layered approach, as shown in Figure 1. It consists of the communication and data aggregation layer, the data fusion layer, the analysis layer and the applications layer.

The basis of the architecture is the communication and data aggregation layer. This layer is responsible for the sensors registration and data aggregation functionalities, taking into account that by the term sensors we refer to IoT nodes and smartphones (including mechanisms for collecting crowdsensing data). The FIWARE Orion Context Broker [6] is used as the main component for supporting IoT nodes registration and configuration, while Comet [7] is used as a component able to manage (storing and retrieving) historical context information as raw and aggregated time series context information. With regards to data collection from mobile applications, interfaces for exchange (storing and retrieving) of data with the ENTROPY platform are provided. Real-time and aggregated data monitoring streams may be enabled by ENTROPY platform end users, activating the data aggregation processes and the storage of the data in the ENTROPY horizontally scalable database (instantiation based on MongoDB).

Following, the data fusion layer is responsible for mapping the collected data in well-defined semantic models and making it available for consumption by a set of services. In order to support unified representation of the data and, thus, facilitate their usage in a common way by application developers and researchers, two semantic models are defined, namely the IoT-based energy management semantic model and the behavioural semantic model [5]. In addition to the unified way of data representation and their re-usability, the semantic enrichment of the collected data augments the expressivity of the information and makes possible the realization of semantic reasoning upon them. It should be noted that, upon the mapping of the data based on the defined semantic models, they are stored in JSON-LD format in the ENTROPY database.

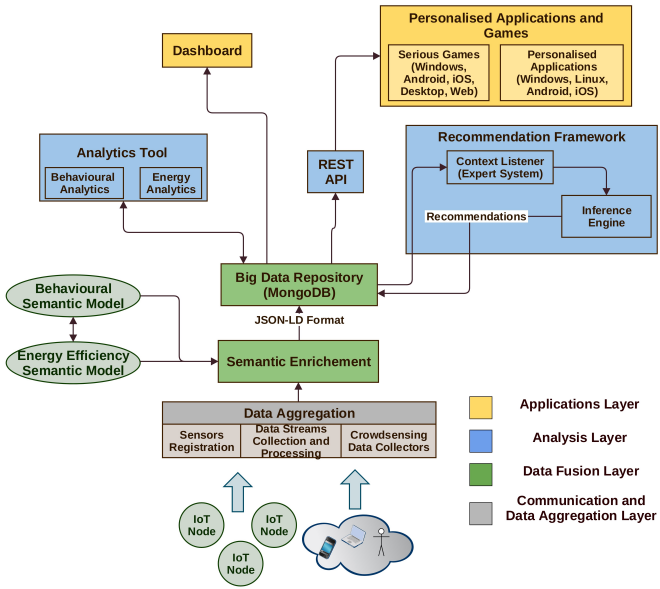

Fig. 1. Conceptual Architecture.

Following, the analysis layer resides on top of the data fusion layer and supports a set of data mining, analysis and personalised recommendation services. As aforementioned, such services consume semantically-mapped data as they are made available by the data fusion layer. The data mining and analysis services provide behavioural and energy analytics. The first ones can lead to insights regarding energy consumption patterns per person and thus to personalised recommendations for behavioural change. The latter ones help building administrators better understand the habits, patterns and preferences of the consumers as well as detect the positive-negative-neutral effect of the gaming and recommendation components upon the behaviour of the consumers. The recommendation mechanisms support the triggering of personalised recommendations, based on rules defined in a rules-based management system.

Lastly, the application layer regards the design and development of mobile applications and serious games that augment energy efficiency awareness of the end users. Through the design and deployment of a set of open APIs as well as the access to unified -in terms of representation- data, the design and development of mobile applications is facilitated through the platform.

### A. Data Aggregation Mechanisms

The Data Aggregation Component is responsible for managing the IoT nodes (sensors) deployed in the buildings as part of the infrastructure. It comprises of different elements, as they are shown in Figure 2 and detailed in the following subsections.

*1) IoT Data Broker*

The IoT Data Broker is the key element of the whole component. In particular, it is in charge of the storage of the historical data from the target IoT nodes. Furthermore, it also keeps the management details of each of these nodes.

For its development, the Orion Context Broker [6] is being used. This broker is part of the FIWARE architecture that provides storage capabilities and a lightweight interface to define and update data entities based on NGSI9-10. In that sense, it makes use of a Non-SQL mongoDB as underlying database and a RESTful interface to access it. As a result, it is a suitable solution when it is necessary to keep the current state of a set of entities of interest, in this case, the IoT nodes. Furthermore, in conjunction with the FIWARE enabler COMET [7] it is capable to store the historic of the measurements returned by the nodes.

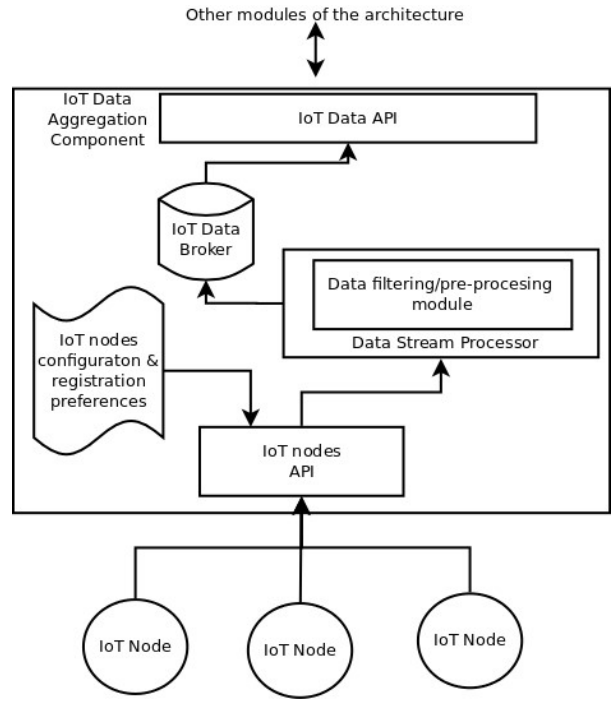

Fig. 2. Data Aggregation Mechanisms Structure.

*2) IoT nodes API*

The IoT nodes API centralizes all the direct access to the IoT nodes or sensors. It delivers the appropriate sequence of commands to these nodes during their bootstrapping stage for their proper configuration. For that goal, it relies to a palette of policies. Once the IoT nodes are running, this module also supports the automatic connection and disconnection of the nodes to the platform in real time. In addition to that, it is the element that directly receives the raw sensor measurements from the nodes. However, the processing of this data is carried out by the Data Stream Collection as we will see later.

We leverage the FIWARE IoT Agent enabler to develop this component. This enabler allows the direct connection with hardware devices so as configure them, check their status and receive their measurements. More specifically, this IoT agent supports several communication protocols to connect with resource-constrained electronic devices, namely COAP MQTT and Lightweight M2M (LWM2M). Moreover, it can be easily connected with the IoT data broker by means of the NGSI interface.

*3) Data Stream Processor*

The Data Stream Processor is in charge of the pre-processing and cleaning of the raw measurements coming from the IoT nodes. To do so, the IoT nodes API redirects all the measurements to this module.

It is developed by following the Complex Event Processing (CEP) approach [8]. CEP is a software paradigm to come up with real time solutions. In a nutshell, a CEP system comprises a set of reactive rules in charge of detecting certain situations of interest by means of the correlation, aggregation and pattern matching over a set of data streams.

Hence, a CEP approach is used in order to detect and remove outliers of the collected measurements. Therefore, it is possible to carry out most of the data pre-processing stage in almost real time.

*4) IoT Data API*

Once the data generated by the IoT nodes is stored in the IoT data broker -upon their processing by the data stream processor, the IoT Data API exposes such data to the rest of the components of the architecture and specifically the Data Fusion Layer. To do so, it defines new high-level entities stored in such a broker that are based on the data from the IoT nodes. For example, it can define an entity "room" aggregating the values of all the IoT nodes located in the same room or spatial area within a building. As a result, this sub-component provides a higher-level and uniform access layer to the platform with respect the raw data stored in the broker. Consequently, other modules of the platform access it by means of RESTful APIs.

*B. Data Fusion Mechanisms*

The semantic enrichment component is responsible for mapping the collected information in the defined ENTROPY semantic models, and thus making possible its usage based on common representation schemas. The continuous evolution of the semantic models, in order to be able to map the collected information in specific entities, along with the appropriate categorization of the available information is considered crucial. Specifically, the two models, namely the behavioural semantic model and the IoT-based energy management semantic model, are being used during the mapping process [5].

The produced output of data is in JSON-LD format. JSON-LD is a lightweight Linked Data format that is easy for humans to read and write. It is based on the already successful JSON format and provides a way to help JSON data to interoperate at web-scale. JSON-LD is an ideal data format for programming environments, REST Web services, and unstructured databases such as MongoDB. The produced JSON-LD data is made available in the ENTROPY database and can be used by the analysis layer tools or used for interlinking purposes with available public or private data.

The collected data is stored at the ENTROPY repository and made available for further processing. Such data regard on one hand data coming from the communication and data aggregation layer as well as data collected from the data mining and analysis tools, the recommendation engine or the ENTROPY personalised application and serious game.

The implementation of the repository is based on MongoDB that is a free and open-source cross-platform document-oriented database. Classified as a NoSQL database, MongoDB avoids the traditional table-based relational database structure in favour of JSON-like documents with dynamic schemas, making the integration of data in certain types of applications easier and faster. MongoDB supports a set of scaling and high performance assurance characteristics. A set of collections is implemented for the storage of the data coming from the various components/layers along with the appropriate interfaces for providing access to these data from the associated tools.

*C. Recommendation and Analysis Framework*

ENTROPY platform hosts a recommender engine, that creates recommendations based on context changes defined on sensor data streams. The recommender engine provides personalized recommendations by taking user attributes and behavioural traits into account. These recommendations are then used by the ENTROPY applications, namely the personalized mobile application and serious game.

The behavioural interventions can be done in many ways, for instance constant feedback to consumers about their energy consumption. While providing feedback to consumers is effective, the interventions may increase their impact by personalization and context awareness. A report [9] that investigates the human factor in energy efficiency shows that it is important to be aware of the context, drivers to which the consumer susceptible and the adaptation of an intervention. Therefore, within the ENTROPY platform, we provide a rule based recommendation engine, that personalizes intervention content according to user's behavioural traits. The personalization occurs by matching specific content with certain intervention factors.

The implementation of the recommendation engine consists of two main steps: (a) creation of rules that match predefined user groups with customized recommendation content, based on the surveys done with the pilot sites; and (b) discovery of new group memberships for users based on the feedback they give to the initial recommendations.

The recommender engine uses condition-action rules that uses context changes as condition and initiates the business logic that selects target users for recommendations based on the matching that is made in the action part of the rule.

An abstract representation of a rule is shown in Table I. The first requirement for the rule is the detection of the context change, which is the change in the measurement of the $CO_2$ level in a certain building space. The measurements are defined as sensor data streams, which consist of the components like the sensor, the frequency and the type of the measurement. For the example in Table I, the sensor data stream may be defined with the sensor monitoring the $CO_2$ levels with a data sampling frequency of one hour. The type of the measurement can be the last value measured. This means the rule condition checks every hour, if the last measurement in a building space (e.g. an office) is over 1000 ppm.

TABLE I. PERSONALISED CONTENT BASED ON GAMER TYPE

| Humanitarian | Socialiser | Free Spirit |
|---|---|---|
| The air quality can become better. Let's open the door for 2 minutes to freshen up and get closer to earning the Refresher Badge (after N times of action) | The air quality is poor for all in the office. Open the door for 2 minutes to freshen the atmosphere and become the Fresh Air Challenge team leader for now | The air quality is quite poor. Open the door for 2 minutes to freshen up and get closer to unlocking a new functionality (N more actions remaining) // to progress to the next level (N more actions remaining) |

When the condition holds, the recommender engine assigns target users to the selected recommendations. The target user groups are defined by the system administrator. An example group may be the Humanitarian users who have activities at the office where the $CO_2$ sensor is located.

Every recommendation receives a feedback from the user, although the nature of the feedback may vary. Initially, we distinguish two types of recommendations, namely, Task and Message. The difference comes with the nature of feedback and the validation method. Tasks require one or a series of action from the user that can be validated by the existing infrastructure. If we go back to the $CO_2$ sensor scenario, in case the door has a sensor, the platform can validate whether the user took the recommended action and whether it had an effect. The validation method can be derived from the recommendation rule, in terms of involved building objects (e.g. door) and sensors. Contrary to tasks, messages cannot be validated. These are, for instance, simple tips to raise awareness of a topic or yes/no questions to collect further information about user's preferences. In this case, the platform accepts users' feedback by default without any validation.

Other type of considered recommendations regard quizzes provided to end users for increasing their energy-efficiency awareness, as well as messages in the form of questions for collection crowdsensing-based feedback based on provided question.

All the terms and the relationships between that recommendation engine uses to understand the data are defined in Entropy semantic models [5]. The recommendation engine makes a heavy use of one of these models, namely, ENTROPY Behavioural Intervention Ontology (EBIO) which represents the interventions and user attributes, such as demographics and interests and personality types. In the second step, we will reason over user's behavioural model, in order to infer new group memberships. An example class expression for the reasoning of new group memberships is shown in Listing 1.

| Player *equivalentTo* Person that *hasPreference* some Reward and *hasPreference* some Competition |
|---|

Listing 1. An OWL DL class expression for inferring gamer types in OWL Manchester Syntax.

We can infer that the users who have interest in rewards and competition are in Player type, with the help of the OWL DL class expression in Listing 1.

Also, data analytics techniques are used to provide recommendations based on previous events, predictions of future ones and are involved in the processing and management layers of the data workflow. Behavioural analytics are also supported for classifying users in set of groups (e.g. by applying clustering techniques) and then providing targeted recommendations per group of users. The realization of analysis is based on the definition of templates for the support of an algorithm, the provision of a set of configuration options and the design or selection of a query for fetching the input data to be used for the analysis.

III. ENTROPY PLATFORM DEVELOPMENT AND APPLICATION AT USE CASES

The ENTROPY platform regards the main interface for providing the set of designed services (data mining and analysis, visualisations, recommendations) where integration activities of the aforementioned mechanisms have taken place. Upon the registration of the set of building spaces where an energy efficiency campaign is going to take place and the specification of the main configuration parameters, the set of registered sensors are made available in the platform for data collection and management purposes.

Following, a set of sensor data streams can be activated. Such streams regard real-time monitoring data streams as well as aggregated data streams (e.g. get avg, min or max values from historic data for predefined time periods). The set of activated sensor data streams may be consumed by the platform itself for visualasation, data analysis or recommendation purposes, or by the developed personalized applications and serious games for supporting the specified sensor data management functionalities. An indicative visualization is provided at Figure 3.

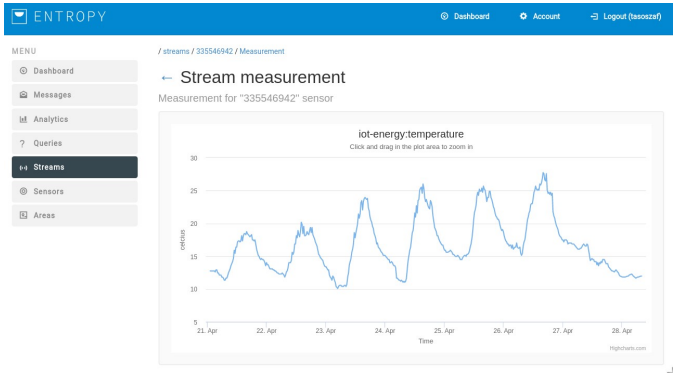

Fig. 3. Sensor Data Stream Visualisation.

Given the collection of sensor data streams and the storage of the semantically mapped data in the ENTROPY repository, their consumption via the set of services can be realized. In order to facilitate the preparation of the suitable datasets per case, a query builder for getting data over the MongoDB ENTROPY repository is implemented. The main functionalities supported regard the specification of queries for getting group of users based on user-defined parameters as well as the specification of queries for getting sensor data that can be provided as input for analysis purposes (see Figure 4).

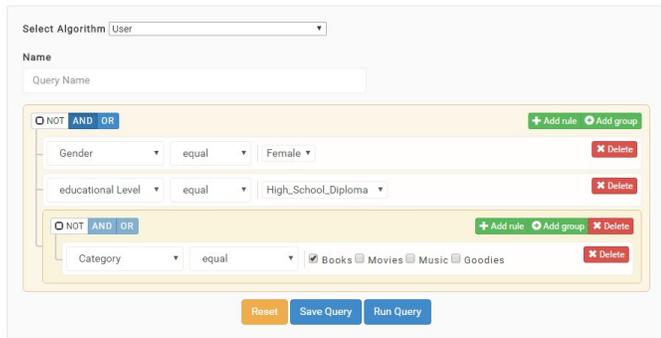

Fig. 4. Query Builder.

A main dashboard is also made available for providing main statistics and informational data with regards to energy efficiency achieved in the registered building (see Figure 5). The main objective of the dashboard is to provide a quick overview of the current status of energy consumption as well as comparisons with previous time periods.

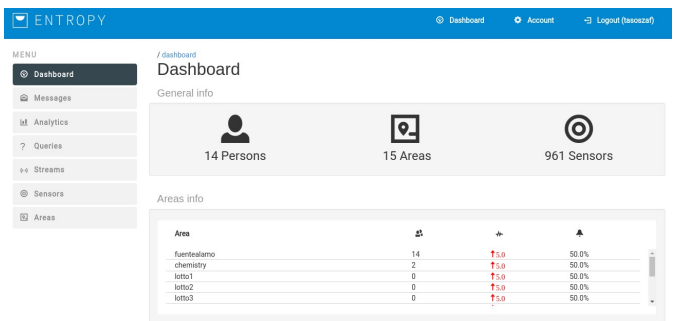

Fig. 5. ENTROPY Platform Dashboard.

The ENTROPY platform is going to be deployed, validated and evaluated in three use cases, covering diverse types of buildings and usage scenarios. Namely, the use cases concern an implementation at the technological park and the university campus of the University of Murcia in Spain, an implementation at the Technopole in Sierre in Switzerland and an implementation at the Navacchio Technology Park close to Pisa in Italy.

At the University of Murcia, three buildings are considered covering main teaching areas, labs as well as offices. A set of sensors for monitoring of energy consumption, HVAC usages as well as a set of environmental parameters (e.g. temperature, humidity) are installed and registered at the ENTROPY platform. User groups engaging university students as well as researches and admin personnel are going to be considered. Similarly, at the Technopole in Sierre as well at the Navacchio Technology Park energy consumption meters along with sensors monitoring environmental parameters as well as sensors identifying presence of people in the rooms are registered in the platform. The main groups that energy efficiency campaigns are going to be applied regard employees' groups.

## IV. CONCLUSIONS AND FUTURE WORK

In this manuscript, we have presented the architectural approach followed in the ENTROPY project, aiming at enabling the design and implementation of a bucket of services, personalized applications and serious games, able to support occupants behavioural change with regards to energy consumption patterns. Towards this direction, a set of novel ICT technologies are exploited leading to the design, implementation and integration of mechanisms supporting the data aggregation and fusion, semantic representation, data mining and analysis, as well as the provision of personalized recommendations.

The set of designed and implemented mechanisms are integrated in the first release of the ENTROPY platform that is shortly presented. This version of the platform is going to be used for the realization of the three ENTROPY use cases in the selected buildings, while in parallel, the designed mechanisms per layer of the architecture are going to be evolved and integrated in the final version of the platform.

The main objective is to achieve reduction of energy consumption of smart buildings by exploiting on one hand the capabilities provided by novel ICT technologies and on the other hand the set of personalised motives for adopting more energy efficient lifestyles.

ACKNOWLEDGMENT

This work was supported by the European Commission Research Programs through the ENTROPY Project under Contract H2020-649849.

## REFERENCES

[1] Energy efficiency in buildings, European Commission, Available Online: https://ec.europa.eu/energy/en/topics/energy-efficiency/buildings

[2] Smarter2030 report, Available Online: http://smarter2030.gesi.org/

[3] ENTROPY H2020 project, Available Online: http://entropy-project.eu/

[4] Barbu, A.-D., Griffiths, Nigel, & Morton, G. (2013). Achieving energy efficiency through behaviour change: what does it take? Luxembourg: Publications Office of the European Union, 2013. Copenhagen. Retrieved from http://www.engerati.com/sites/default/files/Day2-1440-AncaDianaBarbu-EUW2013.pdf

[5] U. Şimşek, A. Fensel, A. Zafeiropoulos, E. Fotopoulou, P. Liapis, T. Bouras, F. T. Saenz, A. F. Skarmeta Gómez. 2016. A semantic approach towards implementing energy efficient lifestyles through behavioural change. In Proceedings of the 12th International Conference on Semantic Systems (SEMANTiCS 2016), ACM, New York, NY, USA, 173-176. DOI: http://dx.doi.org/10.1145/2993318.2993346

[6] Orion Context Broker Documentation, Available Online: https://fiware-orion.readthedocs.io

[7] COMET Documentation, Available Online: https://fiware-sth-comet.readthedocs.io

[8] O. Etzion, P. Niblett, Event processing in action. Manning Publications Co, 2010.

[9] A. Pegels, A. Figueroa, B. Never, The Human Factor in Energy Efficiency, 2015. Retrieved from https://www.die-gdi.de/uploads/media/The_Human_Factor_in_Energy_Efficiency_FINAL_LOW_RES.pdf